# Import test questions into Moodle LMS

Iryna S. Mintii[1[0000-0003-3586-4311]], Svitlana V. Shokaliuk[1[0000-0003-3774-1729]],
Tetiana A. Vakaliuk[2[0000-0001-6825-4697]], Mykhailo M. Mintii[1[0000-0002-0488-5569]]
and Vladimir N. Soloviev[1[0000-0002-4945-202X]]

[1] Kryvyi Rih State Pedagogical University, 54, Gagarina Ave., Kryvyi Rih, 50086, Ukraine
[2] Zhytomyr Polytechnic State University, 103, Chudnivska Str., Zhytomyr, 10005, Ukraine
`irina.mintiy@kdpu.edu.ua`, `shokalyuk@kdpu.edu.ua`,
`{tetianavakaliuk, mikhail.mintii9, vnsoloviev2016}@gmail.com`

**Abstract.** The purpose of the study is to highlight the theoretical and methodological aspects of preparing the test questions of the most common types in the form of text files for further import into learning management system (LMS) Moodle. The subject of the research is the automated filling of the Moodle LMS test database.

The objectives of the study: to analyze the import files of test questions, their advantages and disadvantages; to develop guidelines for the preparation of test questions of common types in the form of text files for further import into Moodle LMS.

The action algorithms for importing questions and instructions for submitting question files in such formats as Aiken, GIFT, Moodle XML, "True/False" questions, "Multiple Choice" (one of many and many of many), "Matching", with an open answer – "Numerical" or "Short answer" and "Essay" are offered in this article. The formats for submitting questions, examples of its designing and developed questions were demonstrated in view mode in Moodle LMS.

## 1 Introduction

Pedagogical testing, due to its high technological and informative content, has surely become a leading method of research into the structure of educational achievement [4, p. 13]. Evidence of it is the introduction in the system of general secondary education external independent assessment and state final certification [7]. Computer-based testing is considered to be the most standardized and objective method of monitoring and evaluating learning outcomes [9]. Requirements for computer testing include:

1. testing variability;
2. prompt submission of student diagnostic results;
3. prompt processing of test results;
4. application of adaptive testing algorithm;
5. accumulation of test results and analysis of their dynamics;

6. dynamic design of tests [4, p. 18].

Computerized testing at Moodle LMS enables to meet most of these requirements – generating test questions randomly from an existing bank, automatically mixing the order of test questions and answer options (alternatives), having different assessment options ("adaptive mode", "deferred feedback", "immediate feedback", etc.), recording the results of each test attempt at evaluation logs and more.

A considerable number of questions are required to provide meaningful validity for the test. However, developing of such questions in Moodle LMS directly in the browser is time consuming – it takes a lot of time and attention. You can significantly reduce the time for filling a bank of test questions of considerable volume by preparing and importing questions in the form of a text file that corresponds to one of the modern formats for the exchange of test tasks – Aiken format, Blackboard, Embedded answers (Cloze), Examview, GIFT format, Missing word format, Moodle XML format and etc.

The purpose of this article is to highlight the theoretical and methodological aspects of preparing the test questions of the most common types in the form of text files for further import into Moodle LMS.

## 2    Import questions from file

This article examines the peculiarities of preparing for import test questions of the most commonly used types – "True/False", "Multiple Choice" ("one of many" and "many of many"), the question of "Matching", an open-ended question ("Numerical" or "Short answer") or "Essay" in Aiken, GIFT and Moodle XML formats (Fig. 1).

**Fig. 1.** Import page of questions from the file

The Aiken format is extremely simple [1]. However, only "Multiple Choice" questions can be prepared in this format with one correct answer. The detailed algorithm for preparing and importing questions in Aiken format is shown in Table 1.

**Table 1.** The algorithm of actions for import in Aiken format

| Step 1 |
| --- |
| Open the window for any text editor (or processor) to work |

| Step 2 |
| --- |
| Make a list of test questions and answer options consistently (one after another) strictly in the format:<br><br>```<br>The text of the question<br>A. correct answer<br>B. wrong answer 1<br>C. wrong answer 2<br>D. wrong answer 3<br>ANSWER: A<br>```<br><br>*Note.*<br><br>1. The number of alternatives to choose the correct answer cannot exceed 10<br>2. There is no need to waste time choosing the correct answer (variation A, B, C, or D), since in Moodle, mixing or not mixing alternatives is configured and performed automatically on the test options page |

| Step 3 |
| --- |
| Save the file as a text document [*],[**] in Unicode encoding mode (UTF-8)<br><br>*Note.*<br>[*] In text editor Notepad: File → Save → File type: Text documents; Encoding: Unicode (UTF-8)<br>[**] In text processor MS Word: File → Save → File Type: Plain Text; Encoding: Unicode (UTF-8) |

| Step 4 |
| --- |
| In Moodle (on the relevant course page), import the saved file to the bank issues by selecting the format of the Aiken file (Fig. 1):<br>4.1. Control Panel → Bank Issues → Import<br>4.2. File format: Aiken<br>4.3. Import questions from a file: Import → Select file ... → ...<br>4.4. After the message is resolved from the import file and the successful import of all issues is completed, click Continue |

The GIFT format is much more powerful than Aiken, because besides preparing different types of questions ("True/False", "Multiple Choice", "Matching", "Numerical", "Short Answer", "Essay", etc.), it also has the ability to add question names, percentages, graphics, comments [2], and etc.

The detailed algorithm for preparing and importing questions in GIFT format is shown in Table 2.

**Table 2.** The algorithm of actions for import in GIFT format

| Step 1 |
| --- |
| Open the window for any text editor (or processor) to work. |
| **Step 2** |
| Make a list of test questions and answer options according to the sample and instructions in Table 4:<br><br>`The text of the question`<br>`{`<br>`answers`<br>`}`<br><br>or (if necessary, enter the name of the question):<br><br>`:: The title of the question :: The text of the question`<br>`{`<br>`answers`<br>`}` |
| **Step 3** |
| Save the file as a text document in Unicode encoding mode (UTF-8) |
| **Step 4** |
| Import saved file (in case of use of images – archive) to the bank of questions, choosing the format of the file GIFT (Fig. 1) |

Preparing Moodle XML questions is not easy at first sight. An example of a file fragment (resulting from export) is shown in Fig. 2.

The ability to work with an intuitive interface while creating questions of various types (with the addition of images, question names, comments, category creation, etc.) in the MS Word text processor environment necessitates the use of the Moodle Quiz macro (Fig. 3) [6].

The detailed algorithm for preparing and importing questions in Moodle XML format using the Word template with the Moodle Quiz macro is shown in Table 3.

**Table 3.** The algorithm of actions for import in the format of Moodle XML

| Step 1 |
| --- |
| Open the template with the macro moodle_quiz_v21 [56] in the MS Word processor window, if necessary, unlock the macros. For successful execution of actions in the tab of tabs MS Word will appear tab Moodle Quiz (Fig. 3) |
| **Step 2** |
| Make a question using the appropriate tools of the Moodle Quiz tab (see Table 4) |
| **Step 3** |
| Use the tool Check Layout (Fig. 3) to verify the correct test pattern |

| Step 4 |
| --- |
| Use the tool Export to XML (Fig. 3) to export the doc file to the XML format |
| Step 5 |
| Import the saved file to the bank by selecting the format of the Moodle XML file (Fig. 1) |

```xml
<question type="truefalse">
        <questiontext format="html">
                <text>Для створення публікацій використовують MS Publisher?</text>
        </questiontext>
        <image></image>
        <image_base64></image_base64>
        <generalfeedback>
                
        </generalfeedback>
        <penalty>0.1</penalty>
        <hidden>0</hidden>
        <answer fraction="100">
                <text>true</text>
                <feedback>
                        
                </feedback>
        </answer>
        <answer fraction="0">
                <text>false</text>
                <feedback>
                        
                </feedback>
        </answer>
        <name>
                <text>Питання Так/Ні – відповідь Так</text>
        </name>
</question>
```

**Fig. 2.** The fragment of the file in Moodle XML format

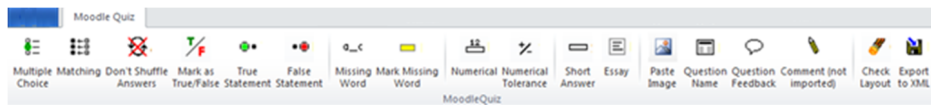

**Fig. 3.** The Moodle Quiz tab

Table 4 provides standards (protocols) and examples of processing different types of questions in text files-documents for importing test questions in GIFT and Moodle XML formats.

**Table 4.** Instructions for submitting questions files in GIFT and Moodle XML formats

| **GIFT format** | **Moodle XML Format (using Moodle Quiz in MS Word)** |
| --- | --- |
| The question "True/False" (Fig. 4) | |
| Format:<br><br>`Question {TRUE}`<br><br>or else<br><br>`Question {FALSE}` | The tool True Statement (Fig. 3) – for the answer to the question Yes. And False Statement (Fig. 3) – for the answer to the question No. |

| GIFT format | Moodle XML Format (using Moodle Quiz in MS Word) |
|---|---|
| `Question Yes/No?`<br>`{TRUE}` | |

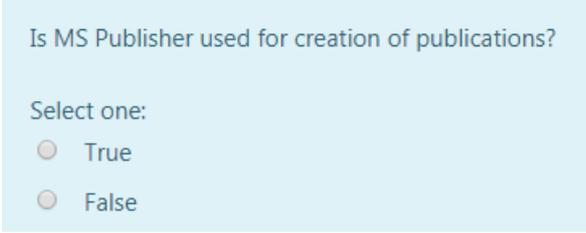

**Fig. 4.** Example of the question "True/False"

The question "Multiple Choice" (Fig. 5, 6)

| Format:<br><br>`Question {= ~ ~~}`<br><br>Example:<br><br>`The question with one correct answer?`<br>`{`<br>`= The correct answer`<br>`~ Wrong answer 1`<br>`~ Wrong answer 2`<br>`~ Wrong answer 3`<br>`}`<br><br>Format:<br><br>`Question {~% number% ~% number% ~}`<br><br>`The questions with several correct answers?`<br>`{`<br>`~% 50% Correct answer 1`<br>`~% 50% Correct answer 2`<br>`~% -50% Wrong answer 1`<br>`~% -50% Wrong answer 2`<br>`}` | The tool Multiple Choice (Fig. 3)<br><br>`The question with one correct answer?`<br>`Correct answer`<br>`Wrong answer 1`<br>`Wrong answer 2`<br>`Wrong answer 3`<br><br>`The questions with several correct answers?`<br>`Correct answer 1`<br>`Correct answer 2`<br>`Wrong answer 1`<br>`Wrong answer 2`<br><br>Note: You can see the answer to the opposite (from right to wrong) using the tool Mark as True/False (Fig. 3). |

| GIFT format | Moodle XML Format (using Moodle Quiz in MS Word) |
|---|---|
| Note: if there are three correct answers to the question, then each of them should add %33.333%, if four – % 25%, etc. | |

What is the extension of text documents created in MS Word?

Select one:
- ○ doc, docx
- ○ ppt, pptx
- ○ htm, html
- ○ jpeg, tiff

**Fig. 5.** Example of the question "Multiple Choice" (one of many)

Choose Google services from the following:

Select one or more:
- ☐ Disc
- ☐ Gmail
- ☐ Youtube
- ☐ Word
- ☐ Skype

**Fig. 6.** Example of the question "Multiple Choice" (many of many)

| The question "Matching" (Fig. 7) | |
|---|---|
| Format:<br><br>```<br>Question {= Questions -> Answer}<br><br>Questions about matching:<br>{<br>= Question 1 -> Answer 1<br>= Question 2 -> Answer 2<br>= Question 3 -> Answer 3<br>= Question 4 -> Answer 4<br>= -> Answer 5<br>}<br>``` | The tool is Matching (Fig. 3). Pressing the Enter key means the beginning of the introduction of question 1, pressing Enter again – the beginning of the input Answers 1, etc.<br>In the end, leave one question blank and enter an additional answer.<br><br>```<br>Questions about matching:<br>Question 1<br>``` |

| **GIFT format** | **Moodle XML Format (using Moodle Quiz in MS Word)** |
|---|---|
| | ```
Answer 1
Question 2
Answer 2
Question 3
Answer 3
Question 4
Answer 4

Answer 5
``` |

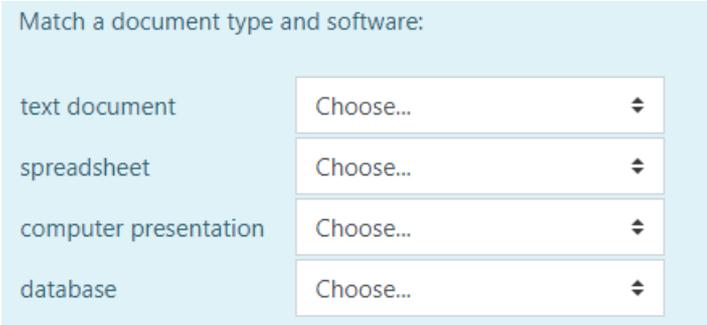

**Fig. 7.** Example of the question "Matching"

| The question "Numerical" (Fig. 8) | |
|---|---|
| Format 1:<br><br>```Question {# number}```<br><br>Format 2:<br><br>```Question {#min value..max value}```<br><br>```Numerical question 2 + 2?```<br>```{# 4}``` | The tool Numerical (Fig. 3). To enter an answer – Enter. To enter accuracy – the tool Numerical Tolerance (Fig. 3).<br><br>```Numerical question 2 + 2?```<br>```4``` |

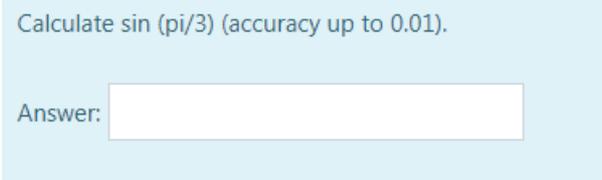

**Fig. 8.** Example of the question "Numerical"



| **GIFT format** | **Moodle XML Format (using Moodle Quiz in MS Word)** |
|---|---|
| The question "Short Answer" (Fig. 9) ||
| Format:<br><br>`Question {= answer}`<br><br>`The question with a short answer?`<br>`{`<br>`= yes`<br>`}` | The tool Short Answer (Fig. 3). Pressing the Enter key means entering the answer.<br><br>`The question with a short answer?`<br>`Yes` |

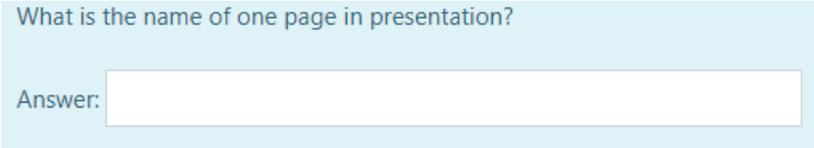

**Fig. 9.** Example of the question "Short Answer"

| The question "Essay" (Fig. 10) ||
|---|---|
| Format:<br><br>`Question {}`<br><br>Example:<br><br>`Task - essay.`<br>`{`<br>`}` | The tool Essay (Fig. 3).<br><br>`Task-essay` |

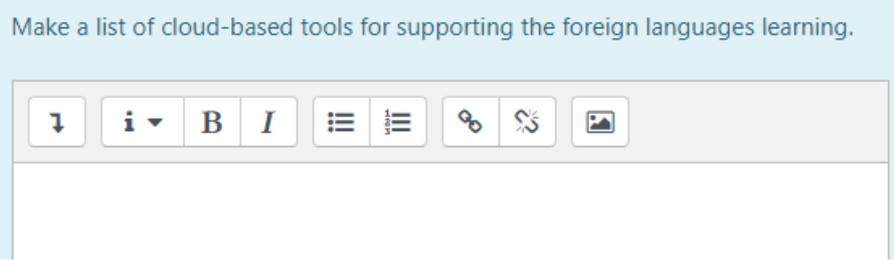

**Fig. 10.** Example of the question "Essay"

| Adding images (in the text of the question or answer variants) ||
|---|---|
| 1. All the images used in this file are saved in the folder (case sensitive) | Tool Paste Image (Fig. 3) (pre-copy the image to the clipboard) |



| GIFT format | Moodle XML Format (using Moodle Quiz in MS Word) |
|---|---|
| 2. Place the `` tag on the image, where name is the name of the image<br>3. When you finish editing, create a zip archive containing the folder and the file with the questions<br>4. The format for importing questions in Moodle LMS – GIFT with medials format (choose zip-archive) | |

Note (for GIFT files).

1. Questions are separated by an empty line, the question itself can not contain empty lines.
2. The text of the question should not contain special characters ({,}, =, ~, #) since they divide the parts of the question. If necessary, they must be preceded by the symbol "\" before each of these characters. It will be deleted when it is imported.
3. If it is necessary to write certain explanations for test users, developers can write a comment starting with the characters "//". The starting point for commenting on answer options is the "#" character.
4. Formatting the text of questions or variants:

```
[html] <p> Questions about formatting </ p>
{
}
```

The main tags for formatting are given in Table 5.

**Table 5.** Tags for formatting text (GIFT format, [8])

| Syntax | Action |
|---|---|
| `<h1> Text </h1>` | heading 1 level |
| `<p> Text </p>` | text paragraph |
| `<br>` | new line |
| `<hr>` | horizontal line |
| `<b> Text </b>` | bold text |
| `<i> Text </i>` | text outline in italics |
| `<sub> Text </sub>` | lower index |
| `<sup> Text </sup>` | top index |
| `<ol>`<br>`<li> List item 1</li>`<br>`<li> List item 2</li>` | numbered list |



| Syntax | Action |
|---|---|
| `<li>` ... `</li>`<br>`</ol>` | |
| `<ul>`<br>`<li>` List item 1 `</li>`<br>`<li>` List item 2 `</li>`<br>`<li>` ... `</li>`<br>`</ul>` | marked list |
| `<a href="URL-link" >` hyperlink text `</a>` | hyperlinks |

## 3 Conclusions

The choice of file format for importing questions depends on the needs of the test developer, and may vary depending on the situation (Table 6).

**Table 6.** Compare file characteristics for importing issues

| Characteristic \ Format | Aiken | GIFT | Moodle XML (macro Moodle Quiz) |
|---|---|---|---|
| Minimalistic interface | + | + | − |
| Different types of questions | − | + | + |
| Images, sounds | − | +<br>(GIFT with media format) | + |
| Automatically formatting | − | − | + |
| Free software | + | + | − |

Yes, the undeniable advantage of the Aiken format is its simplicity, but the questions prepared in this format are the same. The GIFT format, like Moodle XML, provides the ability to fill questions with different types of questions; however, in GIFT format, all tags should be manually written. The downside of the moodle_quiz_v_21 macro is development for commercial software – MS Word.